\documentclass[aps,prd,10pt,amsmath,amsfonts,twocolumn,nofootinbib,notitlepage]{revtex4-1}

 \usepackage[OT4]{fontenc}

\usepackage{bm}
\usepackage{graphicx}
\usepackage{physics}

\usepackage{braket}
\usepackage{amssymb}
\usepackage{dsfont}
\usepackage{bm}
\usepackage{amsmath}
\usepackage{graphicx}
\usepackage{geometry}

\usepackage{color}

\usepackage{physics}
\usepackage[colorlinks]{hyperref}

\hypersetup{
	bookmarksnumbered,
	pdfstartview={FitH},
	citecolor={green},
	linkcolor={darkblue},
	urlcolor={darkblue},
	pdfpagemode={UseOutlines}}
\definecolor{darkgreen}{RGB}{20,100,20}
\definecolor{darkblue}{RGB}{0,0,130}
\definecolor{darkred}{rgb}{.8,0,0}

\begin{document}

\title{Multimode theory of Gaussian states in uniformly accelerated frames}

\author{Kacper D\k{e}bski}
\email{k.debski@student.uw.edu.pl}
\affiliation{Institute of Theoretical Physics, Faculty of Physics, University of Warsaw, Pasteura 5, 02-093 Warsaw, Poland}

\author{Andrzej Dragan}
\email{dragan@fuw.edu.pl}
\affiliation{Institute of Theoretical Physics, Faculty of Physics, University of Warsaw, Pasteura 5, 02-093 Warsaw, Poland}

\date{\today}

\begin{abstract}
We use the formalism of noisy Gaussian channels to derive explicit transformation laws describing how an arbitrary multimode Gaussian state of a scalar quantum field is perceived by a number of accelerating observers, each having access to at least one of the modes. Our work, which generalizes earlier results of Ahmadi, {\em et al.}, Phys. Rev. D {\bf 93}, 124031 (2016), is the next step torwards a better understanding of the effect of gravity on the states of quantum fields.
\end{abstract}

\maketitle

\section{Introduction}
The paradigm of quantum field theory offers the most accurate description of reality on a microscopical level we currently have. It takes into account all known types of fundamental particles and their mutual interactions. Among these interactions, gravity has a special status \cite{Birrell}. While all the other types are represented by quantum fields, gravity is taken into account into the theory, as a classical curved playground for the quantum dynamics of all the other fields \cite{Ball}. Moreover, the effect of gravity is believed to be locally equivalent to the effect of non-inertial motion of the observer. And as such can still lead to non-trivial consequences in the dynamics of quantum fields including decay rates of unstable particles \cite{Lorek}. Therefore a simpler approach to the effect of gravity on quantum fields has been developed. The one that considers the perspective of uniformly accelerated observers and its relation to the observations of inertial observers \cite{Fuentes}. Since the discovery of the Unruh effect, it has been known that a transformation between different observers involves a non-trivial transformation of the state of any quantum field. This has far-reaching consequences to the theory of quantum information and leads to its relativistic generalization that takes into account mutual motion of observers performing quantum-informational protocols such as teleportation \cite{Friis}.

One of the major problems in this approach lies in the difficulty of explicitly writing the transformation laws for a generic state of a quantum field. For a long time the only examples of quantum states that were given an explicit description in the uniformly accelerated reference frames were the vacuum state as well as some simple states defined as excitations of so-called Unruh modes \cite{Bruschi}. Unfortunately, the latter ones were found to be unphysical and therefore no solid description of quantum states other than vacuum was known \cite{Doukas}. A recent progress in this field has been triggered by an observation that it is possible to provide a relatively simple transformation law of Gaussian states between an inertial frame and a uniformly accelerated frame of reference \cite{Dragan1}. This resulted in an immediate application of the finding to the study of degradation of continuous-variable entanglement due to acceleration \cite{Dragan2} as well as continuous-variable teleportation and dense-coding protocols carried out between an inertial and a non-inertial observer \cite{Grochowski1}. The transformation laws have been generalized to an arbitrary two-mode Gaussian state using the language of quantum Gaussian channels \cite{Ahmadi}. In the present work we build on the previous results and present a generic scheme for transforming an arbitrary, multimode Gaussian state to a number of accelerating frames, where each of the modes can be observed by a different accelerated observer. We also show how our scheme can be applied to the case, in which each of the observers accelerates in a different direction in space \cite{Grochowski2}.

The paper is organized as follows. In Sec.~\ref{SecState} we introduce our formalism of noisy channels representing a change of the reference frame, in
Sec.~\ref{SecSol} we discuss Bogolyubov transformations between different decompositions of the field operator, in
Sec.~\ref{SecChara} we provide a complete characterization of a general, multimode Gaussian channel and in
Sec.~\ref{SecExa} we give an example of how our formalism can be used in practice. Finally, Sec.~\ref{SecConcl} concludes this paper.

\section{State transformation as a noisy channel\label{SecState}}
The focus of this paper is on the real scalar massive quantum field in $1+1$ dimensional spacetime. Such a field satisfies the Klein-Gordon equation, $\left(\Box+m^2\right)\phi=0$, written in natural units of $c=\hbar=1$. The field equation can be solved in an arbitrary coordinate system, with the only restriction that the chosen coordinates should allow one for the decomposition of the field operator into positive and negative frequency solutions. We will investigate two such systems - Minkowski coordinates corresponding to an inertial observer, and Rindler coordinates representing uniformly accelerated observers. Consider two alternative decompositions of the field operator $\hat{\Phi}$ involving orthonormal positive frequency solutions in the Minkowski frame, $\phi_i$, and orthonormal positive frequency solutions in the Rindler frame, $\psi_i$:
\begin{align}
\label{packetdecomp}
\hat{\Phi}
=\sum_{i} \phi_i \hat{f}_i+\phi_i^* \hat{f}_i^{\dagger}=\sum_{i} \psi_i \hat{d}_i+\psi_n^* \hat{d}_i^{\dagger}, \end{align}
with corresponding annihilation operators $\hat{f}_i$, and $\hat{d}_i$, respectively. These two families of solutions are labelled by a discrete index $i$, because we will focus on scenarios in which these orthonormal solutions consist of a countable number of wavepackets rather than a continuum of plane-waves.

\begin{figure}
\centering
\includegraphics[width=1\linewidth]{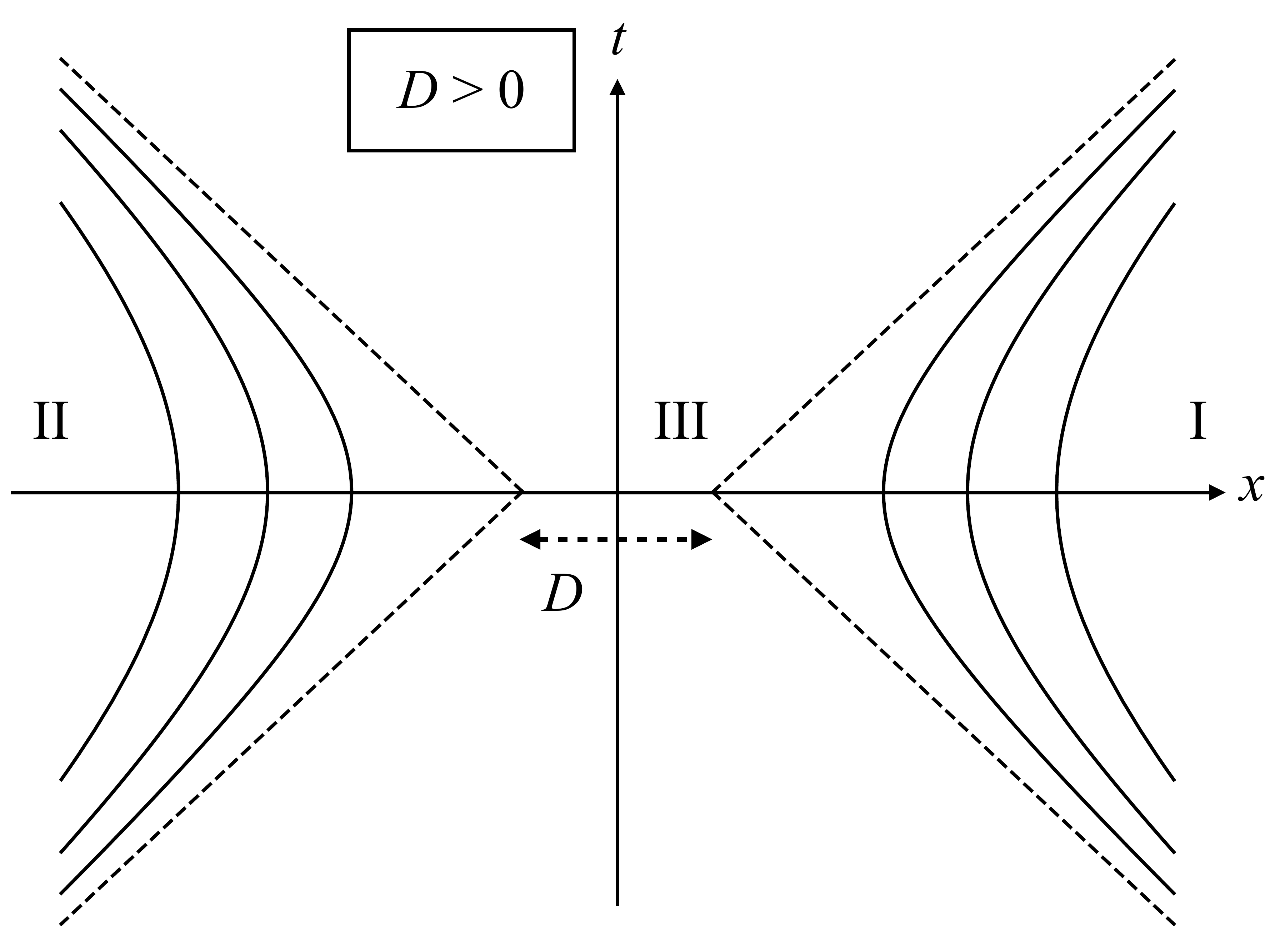}
\includegraphics[width=1\linewidth]{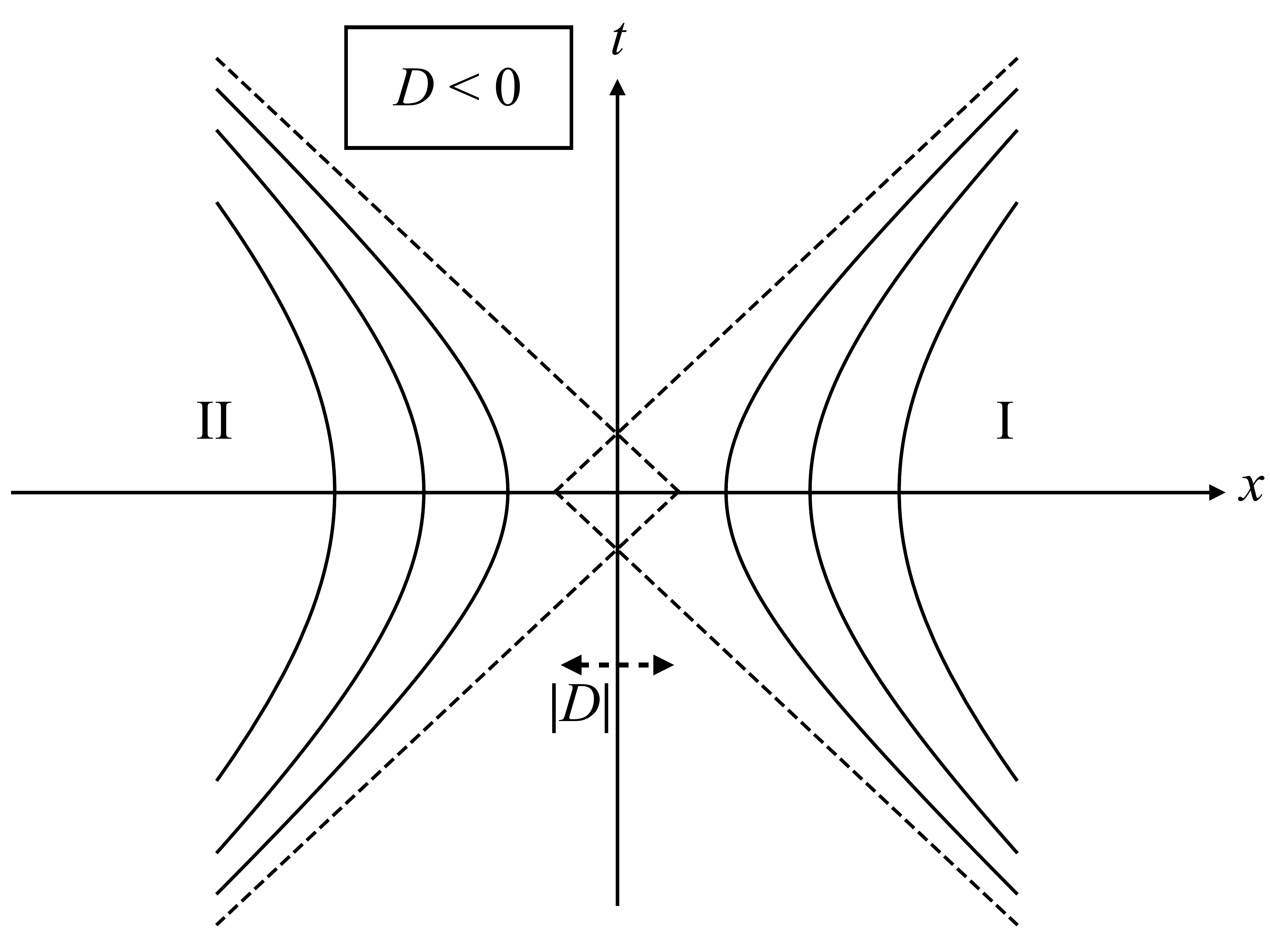}
\caption{Rindler wedges $\mathrm{I}$ and $\mathrm{II}$ in Minkowski spacetime.} 
\label{fig:schemat_stozkow}
\end{figure}

Out of these two infinite families of mode solutions we will select two finite subsets $\{ \phi_n\}_{n\in\left(1,Z\right)}$ and $\{ \psi_n\}_{n\in\left(1,Z\right)}$. We will choose them in such a way that some of these modes will be localized within the Rindler I wedge, and the remaining modes will be localized within the Rindler II wedge, depicted in Fig.~\ref{fig:schemat_stozkow}. Following the construction introduced in \cite{Ahmadi}, we will consider a generalized scenario in which the two wedges are separated by an arbitrary (positive or negative) distance $D$.

Since each of the modes $\phi_n$ consists only of positive Minkowski frequency modes, and each of the modes $\psi_n$ contains only positive Rindler frequencies, and they form orthonormal families of solutions, they must satisfy the following conditions:

\begin{align}
&\forall_{n\neq k}~~ \left(\phi_n\middle|\phi_k^{(*)}\right)=\left(\psi_n\middle|\psi_k^{(*)}\right)=0\mbox{,}
\end{align} 
where the asterisk symbol in the bracket is optional. Similarly, creation and annihilation operators associated with these modes must satisfy the following conditions:
\begin{align}
&\forall_{n\neq k}~~
\left[ \hat{f}_{n},\hat{f}_{k}^{(\dagger)} \right]=
\left[ \hat{d}_{n},\hat{d}_{k}^{(\dagger)} \right]=
0
\mbox{.}
\end{align}
Our goal is to study how quantum information encoded in $Z$ modes $\phi_n$ by an inertial observer can be decoded by an accelerated observer who has only access to $Z$ modes $\psi_n$ defined in his frame. We will assume that the modes of the accelerated observer are chosen such, that each of them corresponds to a single mode of the inertial observer and does not overlap with other ones. Therefore we will impose the following additional condition:
\begin{align}
&\forall_{n\neq k}~~
\left[ \hat{f}_{n},\hat{d}_{k}^{(\dagger)} \right]= 0.
\end{align}
A transition between two reference frames in which a quantum state is defined is described by a linear Bogolyubov transformation of creation and annihilation operators \cite{Birrell}. It has been noticed that such an operation transforms Gaussian states into Gaussian states, which are completely characterized by first and second moments of the quadrature operators defined below. Moreover, the whole operation of changing the reference frame can be described as an action of a quantum Gaussian channel \cite{Ahmadi} on the input states prepared by the inertial observer. The output of such a channel is a Gaussian state observed in the accelerated reference frame. Because of the simplicity of such a description we will be only interested in investigating Gaussian states prepared by the inertial observer. 

The quadrature operators, corresponding to a mode $\hat{f}_k$ are defined as:

\begin{align}
\hat{q}_k^{(f)}\equiv\frac{\hat{f}_{k}+\hat{f}_{k}^{\dagger}}{\sqrt{2}},~~\hat{p}_k^{(f)}\equiv\frac{\hat{f}_{k}-\hat{f}_{k}^{\dagger}}{\sqrt{2}i}
\mbox{.}
\end{align}
Let us define a vector of quadrature operators of all of the $Z$ input modes:

\begin{widetext}
\begin{align}
\hat{\bm{X}}^{(f)}&\equiv\left(\hat{q}_1^{(f)},\hat{p}_1^{(f)},\cdots,\hat{q}_Z^{(f)},\hat{p}_Z^{(f)}\right)^T 
\nonumber
\\
&=\bigg[
\frac{\hat{f}_1+\hat{f}_1^{\dagger}}{\sqrt{2}},\frac{\hat{f}_1-\hat{f}_1^{\dagger}}{\sqrt{2}i},\cdots,\underbrace{\frac{\hat{f}_k+\hat{f}_k^{\dagger}}{\sqrt{2}}}_{\text{2k-1}},\underbrace{\frac{\hat{f}_k-\hat{f}_k^{\dagger}}{\sqrt{2}i}}_{\text{2k}},\cdots,\frac{\hat{f}_Z+\hat{f}_Z^{\dagger}}{\sqrt{2}},\frac{\hat{f}_Z-\hat{f}_Z^{\dagger}}{\sqrt{2}i}
\bigg]^T\mbox{.}
\end{align}
\end{widetext}

The first moment $\bm{X}^{(f)}$ of a quantum state is simply given by an average value of $\hat{\bm{X}}^{(f)}$ \cite{gerardo3}:

\begin{align}\label{mprim5}
\bm{X}^{(f)}=\Big\langle\hat{\bm{X}}^{(f)}\Big\rangle
\mbox{.}
\end{align}
The second moments that form a covariance matrix, are given by \cite{gerardo3}:

\begin{align}\label{mprim6}
\sigma^{(f)}_{kl}\equiv
\Big\langle
\Big\{
\hat{X}_k^{(f)}-X_{k}^{(f)},\hat{X}_l^{(f)}-X_{l}^{(f)}
\Big\}
\Big\rangle,
\end{align}
where $\{\cdot,\cdot\}$ is an anti-commutator. Analogous construction can be carried out for the output modes $\hat{d}_k$ corresponding to the accelerated observers by replacing the letter $f$ with the letter $d$ in the formulae \eqref{mprim5} and \eqref{mprim6}.

The action of a generic Gaussian channel on any Gaussian state be completely characterized by a pair of matrices $M$ and $N$ \cite{10}:
\begin{align}
\label{gausschannel}
\bm{X}^{(d)}&=M \bm{X}^{(f)}\mbox{,}
\nonumber
\\
\sigma^{(d)}&=M\sigma^{(f)}M^T+N\mbox{.}
\end{align}
For $Z$-mode input and output states the $M$ and $N$ matrices are both $2Z\times 2Z$ dimensional. In order to completely characterize the effect of uniform acceleration on the Gaussian states it is sufficient to determine these two matrices. This analysis has been previously carried out for a special case of 2-mode input / output states \cite{Ahmadi}. Here we extend the investigation to account for a generic, multimode Gaussian state.

\section{Solutions to the Klein-Gordon equation\label{SecSol}}
A commonly used orthonormal basis of solutions of the Klein-Gordon equation in an inertial reference frame is given in terms of Minkowski plane waves parameterized by a wavevector $k$:

\begin{align}
u_k=\frac{1}{\sqrt{4\pi \omega_k}}e^{i\left(kx-\omega_k t\right)}
\mbox{,}
\end{align}
where $\omega_k=\sqrt{k^2+m^2}$. The orthonormality conditions can be expressed as: $\left(u_k \middle| u_l \right)=\delta \left( k-l \right)$ and $\left(u_k^* \middle| u_l^* \right)=-\delta \left( k-l \right)$. We will denote annihilation operators associated with these solutions with $\hat{a}_k$.

A similar construction can be carried out in the Rindler reference frame, with the orthonormal solutions corresponding to regions I and II, parameterized by a positive frequency $\Omega$, of the form:

\begin{align}
w_{\text{I}\Omega}&=\sqrt{\frac{\sinh\left(\frac{\pi\Omega}{a}\right)}{\pi^2 a}}K_{i\frac{\Omega}{a}}\left(m\chi\right)e^{-i\Omega\eta}~~ \mbox{in region I}\mbox{,}
\nonumber
\\
w_{\text{II}\Omega}&=\sqrt{\frac{\sinh\left(\frac{\pi\Omega}{a}\right)}{\pi^2 a}}K_{i\frac{\Omega}{a}}\left(-m\chi\right)e^{i\Omega\eta}~~ \mbox{in region II}\mbox{.}
\end{align}
The orthonormality conditions for these solutions take the form: $\left(w_{\text{I}\Omega},w_{\text{I}\Xi}\right)=\delta\left(\Omega-\Xi\right)$, $\left(w_{\text{I}\Omega}^*,w_{\text{I}\Xi}^*\right)=-\delta\left(\Omega-\Xi\right)$, $\left(w_{\text{I}\Omega},w_{\text{I}\Xi}^*\right)=0$, and analogously for region II. In addition, for $D\geqslant 0$ when the regions I and II do not overlap, we have $\left(w_{\text{I}\Omega},w_{\text{II}\Xi}\right)=\left(w_{\text{I}\Omega}^*,w_{\text{II}\Xi}^*\right)=\left(w_{\text{I}\Omega},w_{\text{II}\Xi}^*\right)=0$. The associated annihilation operators corresponding to these solutions will be denoted with $\hat{b}_{\Lambda\Omega}$, where $\Lambda\in\{\text{I}, \text{II} \}$.

All the above annihilation and creation operators satisfy canonical commutation relations. Additionally for $D \geqslant 0$ we have $[\hat{b}_{\text{I}\Omega},\hat{b}_{\text{II}\Omega'}]=[\hat{b}_{\text{I}\Omega}^{\dagger},\hat{b}_{\text{II}\Omega'}^{\dagger}]=[\hat{b}_{\text{I}\Omega},\hat{b}_{\text{II}\Omega'}^{\dagger}]=0$. These conditions are not satisfied for $ D <0 $, due to a non-zero overlap between individual Rindler regions. However, we will still choose all the wavepacket modes such that $\forall_{n\neq k} \left[ \hat{d}_{n},\hat{d}_{k}^{(\dagger)} \right]=0$ for any $D$. 

Since any of the introduced basis is appropriate for the description of the quantum field, we can write the field operator using any of the equivalent decompositions:
\begin{widetext}
\begin{align}\label{eqn:10}
\hat{\Phi}=\int_{-\infty}^{\infty} \text{d}k\, u_k\hat{a}_k+\text{h.c} =\int_{0}^{\infty} \text{d}\Omega \left(w_{\text{I}\Omega}\hat{b}_{\text{I}\Omega}+w_{\text{II}\Omega}\hat{b}_{\text{II}\Omega}\right) + \text{h.c.}+\hat{\Phi}_{\text{III}}\left(D\right)
\mbox{,}
\end{align}
\end{widetext}
where $\hat{\Phi}_{\text{III}}\left(D\right)$ is an additional part of the field operator covering the additional region between regions I and II, labeled as region III. Note that $\hat{\Phi}_{\text{III}}\left(D\right)\neq 0$ even for $D<0$ and vanishes only when $D=0$ \cite{Ahmadi}. All the results discussed in this work are independent of the specific details of $\hat{\Phi}_{\text{III}}\left(D\right)$.

We have introduced two decompositions of the field operator into wavepacket basis of modes \eqref{packetdecomp} and into continuous frequency modes \eqref{eqn:10}. Since the wavepackets contain only positive frequency modes of their respective frames, we have:

\begin{align}
&\forall_{n\in \left(1,Z\right)~k\in \mathbb{R}
\Omega>0} 
\nonumber
\\
&\left(\phi_{n},u_k^*\right)=\left(\psi_{n},w_{\text{I}\Omega}^*\right)=\left(\psi_{n},w_{\text{II}\Omega}^*\right)=0.
\end{align} 
We can also relate the associated annihilation operators via the following identities:

\begin{align}
\hat{d}_{n}&=\int_{0}^{\infty}\mathrm{d}\Omega \left( \left(\psi_{n},w_{\text{I}\Omega}\right)\hat{b}_{\text{I}\Omega}+\left(\psi_{n},w_{\text{II}\Omega}\right)\hat{b}_{\text{II}\Omega}\right),
\\
\hat{f}_{n}&=\int_{-\infty}^{\infty} \mathrm{d}k \left(\phi_{n},u_k\right)\hat{a}_k
\mbox{.}
\end{align}
As a consequence, the canonical commutation relations for the wavepacket annihilation operators lead to the following requirements:
\begin{widetext}
\begin{align}
1=&
\int_{0}^{\infty}\int_{0}^{\infty}
 \mathrm{d}\Omega \mathrm{d}\Omega' \Big(
\left(\psi_{n},w_{\text{I}\Omega}\right)\left(\psi_{n},w_{\text{I}\Omega'}\right)^*[\hat{b}_{\text{I}\Omega},\hat{b}_{\text{I}\Omega'}^{\dagger}]+
\left(\psi_{n},w_{\text{I}\Omega}\right)\left(\psi_{n},w_{\text{II}\Omega'}\right)^*[\hat{b}_{\text{I}\Omega},\hat{b}_{\text{II}\Omega'}^{\dagger}] \nonumber
\\
&+
\left(\psi_{n},w_{\text{II}\Omega}\right)\left(\psi_{n},w_{\text{I}\Omega'}\right)^*[\hat{b}_{\text{II}\Omega},\hat{b}_{\text{I}\Omega'}^{\dagger}]+
\left(\psi_{n},w_{\text{II}\Omega}\right)\left(\psi_{n},w_{\text{II}\Omega'}\right)^*[\hat{b}_{\text{II}\Omega},\hat{b}_{\text{II}\Omega'}^{\dagger}] \Big)
\mbox{,}
\end{align}
\end{widetext}
that restrict the possible choice of the modes. When $D\geqslant0$ then the above condition simplifies to:
\begin{align}
1=\int_{0}^{\infty}\mathrm{d}\Omega\left(|\left(\psi_{n},w_{\text{I}\Omega}\right)|^2+|\left(\psi_{n},w_{\text{II}\Omega}\right)|^2\right)\mbox{.}
\end{align}
And for $D<0$ we assume that the above relation is satisfied by the appropriate choice of the wavepacket modes.

\section{Characterization of the multimode Gaussian channel\label{SecChara}}

Let us proceed with the computation of the matrices $M$ and $N$ characterizing the Gaussian channel \eqref{gausschannel}. A generic transformation describing a transition between two reference frames has the following linear form:

\begin{align}
\label{bogo1}
\hat{d_{k}}&=\sum_{n}\alpha_{k,n}\hat{f_n}-\beta_{k,n}\hat{f_n}^{\dagger} \mbox{,}\\
\hat{d_{k}}^{\dagger}&=\sum_{n}-\beta_{k,n}^* \hat{f_n}+\alpha_{k,n}^*\hat{f_n}^{\dagger} \mbox{,}
\end{align}
where

\begin{align}
\alpha_{i,j}&=\left(\psi_i\middle|\phi_j\right),
\nonumber
\\
\beta_{i,j}&=-\left(\psi_i\middle|\phi_j^*\right).
\end{align}
We can substitute the above relations to the definition of the first moments corresponding to the wavepackets in the accelerated frame:
\begin{widetext}
\begin{align}
X_{2n-1}^{(d)} &= \Bigg\langle\frac{\hat{d_n}+\hat{d_n}^{\dagger}}{\sqrt{2}}\Bigg\rangle=
\frac{1}{\sqrt{2}}\langle\sum_{n'=1}^{Z}\left(\alpha_{n,n'}-\beta_{n,n'}^*\right)\hat{f}_{n'}-
\left(\beta_{n,n'}-\alpha_{n,n'}^*\right)\hat{f}_{n'}^{\dagger}\rangle \nonumber 
\\
&=\sum_{n'=1}^{Z}\operatorname{Re}\left(\alpha_{n,n'}-\beta_{n,n'}\right)X_{2n'-1}^{(f)}-\operatorname{Im}\left(\alpha_{n,n'}+\beta_{n,n'}\right)X_{2n'}^{(f)}, \\
X_{2n}^{(d)} &=\Bigg\langle\frac{\hat{d_n}-\hat{d_n}^{\dagger}}{\sqrt{2}i}\Bigg\rangle =
\frac{-i}{\sqrt{2}}\langle\sum_{n'=1}^{Z}\left(\alpha_{n,n'}+\beta_{n,n'}^*\right)\hat{f}_{n'}-
\left(\beta_{n,n'}+\alpha_{n,n'}^*\right)\hat{f}_{n'}^{\dagger}\rangle \nonumber 
\\
&=\sum_{n'=1}^{Z}\operatorname{Im}\left(\alpha_{n,n'}-\beta_{n,n'}\right)X_{2n'-1}^{(f)}+\operatorname{Re}\left(\alpha_{n,n'}+\beta_{n,n'}\right)
X_{2n'}^{(f)}.
\end{align}
\end{widetext}
The above expressions allow us to determine the form of the matrix $M$ appearing in the transformation properties of the first moments \eqref{gausschannel}. This matrix can be cast in the following block form:
\begin{align}
M=
\left(
\begin{array}{ccc}
 \bm{M}_{1,1} & \cdots  &  \bm{M}_{1,Z} \\
 \vdots  & \ddots & \vdots  \\
  \bm{M}_{Z,1} & \cdots  &  \bm{M}_{Z,Z} \\
\end{array}
\right)
\mbox{,}
\end{align} 
where:

\begin{align}
\label{mij}
\bm{M}_{i,j}=
\left(
\begin{array}{cc}
 \operatorname{Re}\left(\alpha _{i,j}-\beta _{i,j}\right) & -\operatorname{Im}\left(\alpha _{i,j}+\beta _{i,j}\right) \\
 \operatorname{Im}\left(\alpha _{i,j}-\beta _{i,j}\right) & \operatorname{Re}\left(\alpha _{i,j}+\beta _{i,j}\right) \\
\end{array}
\right)
\mbox{.}
\end{align}
Therefore the matrix $M$ is completely characterized by the coefficients of the Bogolyubov transformation \eqref{bogo1}.

In order to compute the matrix $N$ characterizing the transformation properties of the covariance matrix given by \eqref{gausschannel} we first consider the transformation properties of the vacuum state, whose covariance matrix is just an identity, $\sigma_{\text{vac}}^{(f)} = \openone$. This allows us to write the matrix $N$ as \cite{Ahmadi}:  

\begin{align}
\label{nviavac}
N=\sigma_{\text{vac}}^{(d)}-MM^T
\mbox{.}
\end{align}
Therefore in order to determine the matrix $N$ it is sufficient to characterize the properties of the Minkowski vacuum state in the accelerated frame of reference. We proceed with this calculation by considering two special cases: when the two Rindler wedges share a common apex, and when they do not.

By explicit calculation given in details in the Appendix \ref{vaccov} we find the form of covariance matrix of the vacuum state in the Rindler frame for $D=0$. Let us choose the indeces $n\in \big\{1,2,\cdots,Z\big\}$, $k\in \big\{n+1,n+2,\cdots,Z\big\}$. It follows that:
\begin{widetext}
\begin{align}
\left(\sigma_{\text{vac}}^{(d)}\right)_{2n-1,2k-1}
=&
\operatorname{Re}
N_{n,k}^{+}
\mbox{,}
\\
\left(\sigma_{\text{vac}}^{(d)}\right)_{2n-1,2k}
=&
\operatorname{Im}
N_{n,k}^{-}
\mbox{,}
\\
\left(\sigma_{\text{vac}}^{(d)}\right)_{2n,2k-1}
=&
\operatorname{Im}
N_{n,k}^{+}
\mbox{,}
\\
\left(\sigma_{\text{vac}}^{(d)}\right)_{2n,2k}
=&
-
\operatorname{Re}
N_{n,k}^{-}
\mbox{,}
\\
\left(\sigma_{\text{vac}}^{(d)}\right)_{2n,2n-1}=&\left(\sigma_{\text{vac}}^{(d)}\right)_{2n-1,2n}
=
\int_{0}^{\infty}
\mathrm{d}\Omega
\frac{\operatorname{Im}\left[
\left(\psi_{n},w_{\text{I}\Omega}\right)\left(\psi_{n},w_{\text{II}\Omega}\right)\right]}{\sinh\left(\frac{\pi\Omega}{a}\right)}
\mbox{,}
\\
\left(\sigma_{\text{vac}}^{(d)}\right)_{2n,2n}=&
1+
\int_{0}^{\infty}
\mathrm{d}\Omega
\left(
|\left(\psi_{n},w_{\text{I}\Omega}\right)|^2+
|\left(\psi_{n},w_{\text{II}\Omega}\right)|^2
\right)
\frac{e^{-\frac{\pi\Omega}{a}}}{\sinh\left(\frac{\pi\Omega}{a}\right)}
\nonumber
\\
&-
\int_{0}^{\infty}
\mathrm{d}\Omega
\frac{\operatorname{Re}\left[
\left(\psi_{n},w_{\text{I}\Omega}\right)\left(\psi_{n},w_{\text{II}\Omega}\right)\right]
}{\sinh\left(\frac{\pi\Omega}{a}\right)}
\mbox{,}
\\
\left(\sigma_{\text{vac}}^{(d)}\right)_{2n-1,2n-1}=&
1+
\int_{0}^{\infty}
\mathrm{d}\Omega
\left(
|\left(\psi_{n},w_{\text{I}\Omega}\right)|^2+
|\left(\psi_{n},w_{\text{II}\Omega}\right)|^2
\right)
\frac{e^{-\frac{\pi\Omega}{a}}}{\sinh\left(\frac{\pi\Omega}{a}\right)}
\nonumber
\\
&+
\int_{0}^{\infty}
\mathrm{d}\Omega
\frac{\operatorname{Re}\left[
\left(\psi_{n},w_{\text{I}\Omega}\right)\left(\psi_{n},w_{\text{II}\Omega}\right)\right] 
}{\sinh\left(\frac{\pi\Omega}{a}\right)}
\mbox{,}
\end{align}

where:

\begin{align}
N_{n,k}^{\pm}
=&
\int_{0}^{\infty}
\mathrm{d}\Omega
\Big(
\frac{
\left(\psi_{n},w_{\text{I}\Omega}\right)
\left(\psi_{k},w_{\text{II}\Omega}\right)
+
\left(\psi_{n},w_{\text{II}\Omega}\right)
\left(\psi_{k},w_{\text{I}\Omega}\right)
}{\sinh\left(\frac{\pi\Omega}{a}\right)}
\nonumber
\\
&\pm
\frac{
e^{\frac{\pi\Omega}{a}}
\left[
\left(\psi_{n},w_{\text{I}\Omega}\right)
\left(\psi_{k},w_{\text{I}\Omega}\right)^*
+
\left(\psi_{n},w_{\text{II}\Omega}\right)
\left(\psi_{k},w_{\text{II}\Omega}\right)^*
\right]
}{\sinh\left(\frac{\pi\Omega}{a}\right)}
\Big)
\mbox{.}
\end{align}
\end{widetext}

Similarly, when the respective Rindler wedges do not have a common apex, $D\neq 0$, the covariance matrix of the vacuum state can also be computed explicitly - see the details in Appendix \ref{vaccovd}. Let us choose the indices $n\in \big\{1,2,\cdots,Z\big\}$, $k\in \big\{n+1,n+2,\cdots,Z\big\}$. Then we find:
\begin{widetext}
\begin{align}
\left(\sigma_{\text{vac}}^{(d)}\right)_{2n,2n}=& 
1
+
\int_{0}^{\infty}
\mathrm{d}\Omega
\frac{
\left(
|\left(\psi_{n},w_{\text{I}\Omega}\right)|^2
+
|\left(\psi_{n},w_{\text{II}\Omega}\right)|^2
\right)
e^{-\frac{\pi\Omega}{a}}
}{\sinh(\frac{\pi\Omega}{a})}
\nonumber
\\
&+
2
\int_{0}^{\infty} \int_{0}^{\infty}
\mathrm{d}\Omega \mathrm{d}\Omega'
I_3(\Omega,\Omega')
\left[
\left(\psi_{n},w_{\text{I}\Omega}\right)
\left(\psi_{n},w_{\text{II}\Omega'}\right)^*
+
\left(\psi_{n},w_{\text{II}\Omega}\right)
\left(\psi_{n},w_{\text{I}\Omega'}\right)^*
\right]
\nonumber
\\
&-
2\operatorname{Re}
\int_{0}^{\infty}\int_{0}^{\infty}
\mathrm{d}\Omega \mathrm{d}\Omega'
I_{1}(\Omega,\Omega')
\left[
\left(\psi_{n},w_{\text{I}\Omega}\right)\left(\psi_{n},w_{\text{II}\Omega'}\right)
+
\left(\psi_{n},w_{\text{II}\Omega}\right)\left(\psi_{n},w_{\text{I}\Omega'}\right)
\right]
\mbox{,}
\\
\left(\sigma_{\text{vac}}^{(d)}\right)_{2n-1,2n-1}=& 
1
+
\int_{0}^{\infty}
\mathrm{d}\Omega
\frac{
\left(
|\left(\psi_{n},w_{\text{I}\Omega}\right)|^2
+
|\left(\psi_{n},w_{\text{II}\Omega}\right)|^2
\right)
e^{-\frac{\pi\Omega}{a}}
}{\sinh(\frac{\pi\Omega}{a})}\nonumber
\\
&+
2
\int_{0}^{\infty} \int_{0}^{\infty}
\mathrm{d}\Omega \mathrm{d}\Omega'
I_3(\Omega,\Omega')
\left[
\left(\psi_{n},w_{\text{I}\Omega}\right)
\left(\psi_{n},w_{\text{II}\Omega'}\right)^*
+
\left(\psi_{n},w_{\text{II}\Omega}\right)
\left(\psi_{n},w_{\text{I}\Omega'}\right)^*
\right] \nonumber
\\
&+
2\operatorname{Re}
\int_{0}^{\infty}\int_{0}^{\infty}
\mathrm{d}\Omega \mathrm{d}\Omega'
I_{1}(\Omega,\Omega')
\left[
\left(\psi_{n},w_{\text{I}\Omega}\right)\left(\psi_{n},w_{\text{II}\Omega'}\right)
+
\left(\psi_{n},w_{\text{II}\Omega}\right)\left(\psi_{n},w_{\text{I}\Omega'}\right)
\right]
\mbox{,}
\\
\left(\sigma_{\text{vac}}^{(d)}\right)_{2n,2n-1}
=&
2
\operatorname{Im}
\int_{0}^{\infty}\int_{0}^{\infty}
\mathrm{d}\Omega \mathrm{d}\Omega'
I_{1}(\Omega,\Omega')
\left[
\left(\psi_{n},w_{\text{I}\Omega}\right)\left(\psi_{n},w_{\text{II}\Omega'}\right)
+
\left(\psi_{n},w_{\text{II}\Omega}\right)\left(\psi_{n},w_{\text{I}\Omega'}\right)
\right]
\mbox{,}
\end{align}
\begin{align}\label{mprim2}
\left(\sigma_{\text{vac}}^{(d)}\right)_{2n-1,2k-1}
=&
\operatorname{Re} N_{n,k}^{+}(D)
\mbox{,}
\\
\left(\sigma_{\text{vac}}^{(d)}\right)_{2n-1,2k}
=&
\operatorname{Im} N_{n,k}^{-}(D)
\mbox{,}
\\
\left(\sigma_{\text{vac}}^{(d)}\right)_{2n,2k-1}
=&
\operatorname{Im} N_{n,k}^{+}(D)
\mbox{,}
\\
\left(\sigma_{\text{vac}}^{(d)}\right)_{2n,2k}
=&
-\operatorname{Re} N_{n,k}^{-}(D)
\mbox{,}
\end{align}

where:
\begin{align}
I_1(\Omega,\Omega')\equiv&
\frac{
e^{\frac{\pi(\Omega-\Omega')}{2a}(1-\frac{D}{|D|})}
}{
2\pi a
\sqrt{
\sinh\frac{\pi\Omega}{a}
\sinh\frac{\pi\Omega'}{a}
}
}
K_{i\frac{\Omega-\Omega'}{a}}(|mD|)
\mbox{,}
\\
I_3(\Omega,\Omega')\equiv&
\frac{
e^{\frac{\pi(\Omega+\Omega')}{2a}(1-\frac{D}{|D|})}
}{
2\pi a
\sqrt{
\sinh\frac{\pi\Omega}{a}
\sinh\frac{\pi\Omega'}{a}
}
}
K_{i\frac{\Omega+\Omega'}{a}}(|mD|)
\mbox{,}
\\
N_{n,k}^{\pm}(D)
=&
2
\Bigg[
\int\int
\mathrm{d}\Omega \mathrm{d}\Omega'
I_{1}(\Omega,\Omega')
\left[
\left(\psi_{n},w_{\text{I}\Omega}\right)\left(\psi_{k},w_{\text{II}\Omega'}\right)
+
\left(\psi_{n},w_{\text{II}\Omega}\right)\left(\psi_{k},w_{\text{I}\Omega'}\right)
\right]
\nonumber
\\
&\pm
\int
\mathrm{d}\Omega
\frac{
\left(\psi_{n},w_{\text{I}\Omega}\right)
\left(\psi_{k},w_{\text{I}\Omega}\right)^*
+
\left(\psi_{n},w_{\text{II}\Omega}\right)
\left(\psi_{k},w_{\text{II}\Omega}\right)^*
}{1-e^{-\frac{2\pi\Omega}{a}}} 
\nonumber
\\
&\pm
\int \int
\mathrm{d}\Omega \mathrm{d}\Omega'
I_3(\Omega,\Omega')
\left[
\left(\psi_{n},w_{\text{I}\Omega}\right)
\left(\psi_{k},w_{\text{II}\Omega'}\right)^*
+
\left(\psi_{n},w_{\text{II}\Omega}\right)
\left(\psi_{k},w_{\text{I}\Omega'}\right)^*
\right]
\Bigg]
\mbox{.}
\end{align}
\end{widetext}

The explicit form of the $N$ matrix can be calculated via the formula \eqref{nviavac}. The above result provides a complete characterization of a Gaussian channel responsible for the transformation of a generic multimode Gaussian state due to uniformly accelerated motion of the observer. It is a direct generalization of a special case derived in \cite{Ahmadi}. At this stage we have not discussed the possible choice of wavepacket modes $\phi_n$ and $\psi_n$ assuming only, that each of them is fully supported within a single Rindler wedge and contains only positive frequencies in its respective frame. Also, the number of accelerated observers involved in the characterization of the quantum state is arbitrary. It only depends on spatial localization of all the modes. For instance, if each of the modes is localized within the same region of spacetime - in principle their state could be measured by a single accelerated observer having access to a multimode measurement device. On the other hand, if each of the modes is localized in a different region, in general a total number or $Z$ accelerated observers is needed in order to observe the overall state. In the next section we discuss a special case of a $Z$-partite symmetric Gaussian state and apply our scheme to describe its properties in a non-inertial frame.

\section{Application example: $Z$-partite symmetric Gaussian state\label{SecExa}}

\begin{figure}
\centering
\includegraphics[width=1\linewidth]{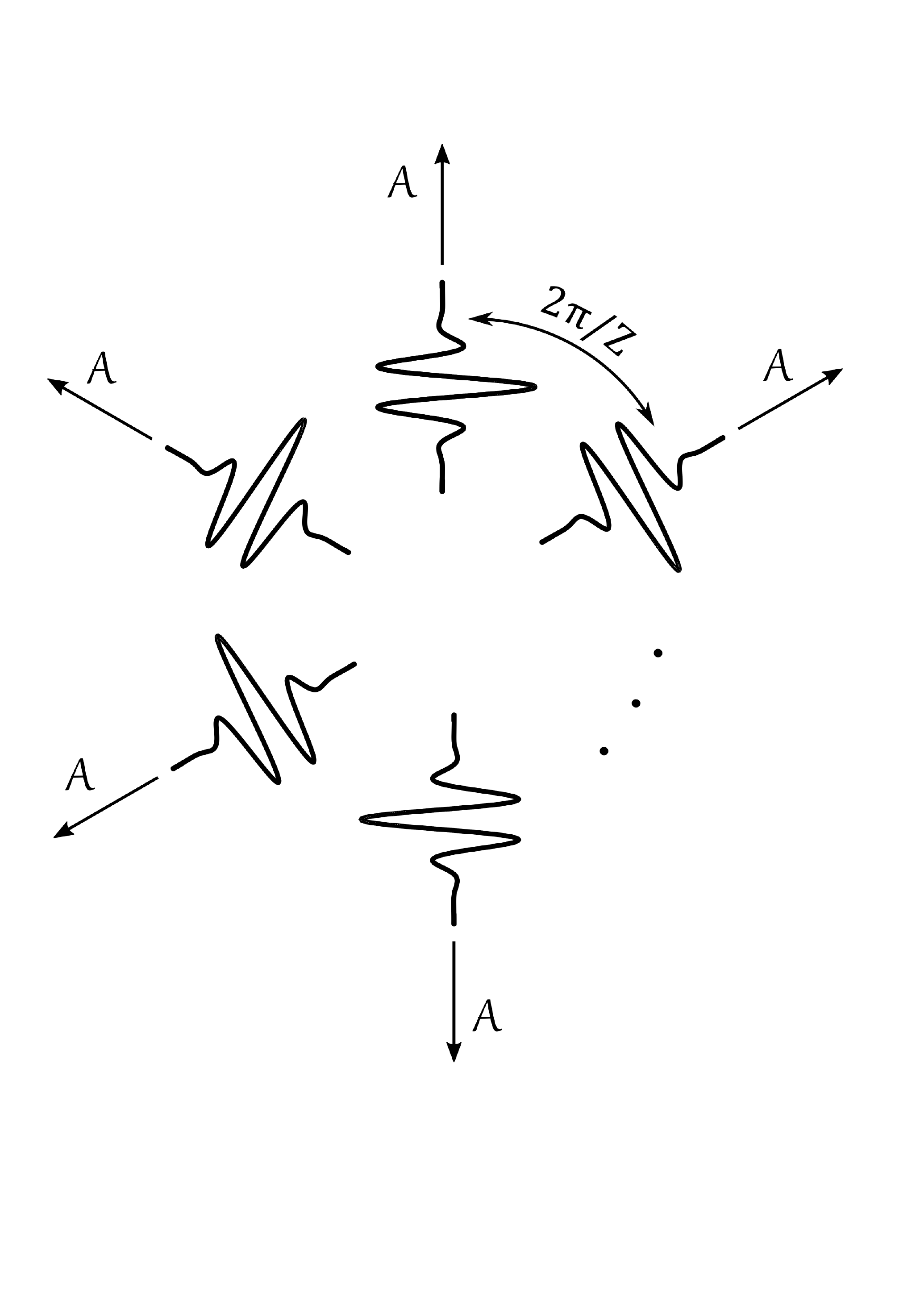}
\caption{Scheme of motion of $Z$ non-inertial observes accelerating with identical proper acceleration $\mathcal{A}$.} 
\label{fig:schemat_obserwatorow}
\end{figure}

Let us apply our multimode quantum channel to the case of non-inertial observers accelerating in $Z$ different directions (see Fig.~\ref{fig:schemat_obserwatorow}). Although the scheme discussed in this work involves only $1+1$ dimensional spacetime, a generalization to the $3+1$ dimensional case discussed in \cite{Grochowski2} allows us to draw conclusions based on our simplified scheme to a good approximation.

\begin{figure}
\centering
\includegraphics[width=1\linewidth]{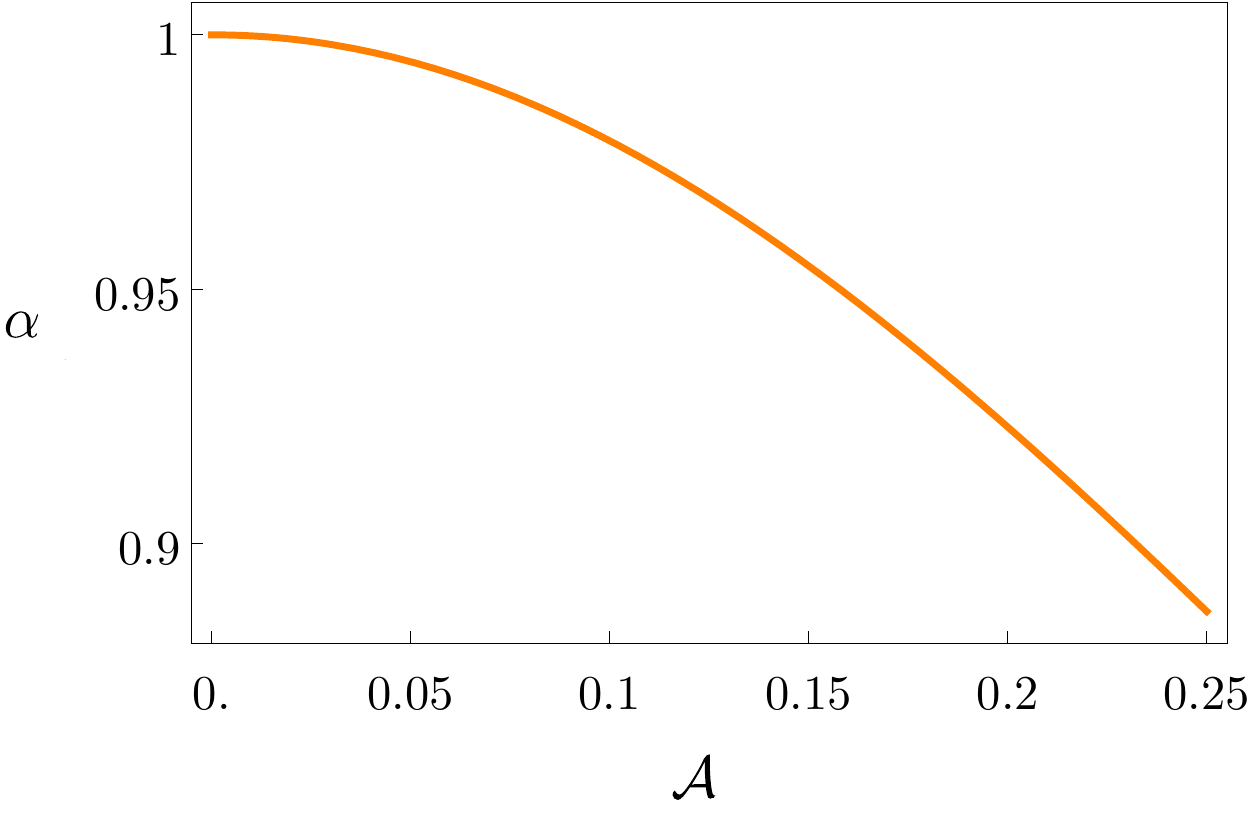}
\caption{The dependence of the $\alpha$ coefficient on the observer's acceleration. Figure from \cite{Grochowski1}.} 
\label{fig:przekrycie_od_a}
\end{figure}

We will consider $Z$ non-inertial observes accelerating with identical proper acceleration. Each of these observers accelerates towards a direction forming a $\frac{2\pi}{Z}$ angle with the remaining pair of accelerated observers, so that the scheme is completely symmetrical. As a consequence, the overlap coefficients appearing in \eqref{mij} will be the same:

\begin{align}
 \alpha_{i,j}=& \delta_{i j}\left(\psi_i\middle|\phi_j\right)
=
\delta_{i j}\alpha
~~\mbox{,}
\nonumber
\\
\beta_{i,j}=&-\delta_{i j}\left(\psi_i\middle|\phi_j^*\right)
=
\delta_{i j}\beta
~~\mbox{.}
\end{align}
In order to proceed with the explicit computation of these coefficients we will choose to work with the mode functions introduced in \cite{Grochowski1}. For these wavepackets the calculated $\beta$ coefficients are at least 8 orders of magnitude smaller than $\alpha$'s and can be safely neglected. For such a choice, the dependence of the $\alpha$ coefficient on the proper acceleration of the observer is shown in Fig.~\ref{fig:przekrycie_od_a}.

The resulting quantum channel can be used for transforming a fully symmetric $Z$-mode pure squeezed vacuum state. The covariance matrix of such state can be written as \cite{gerardo1,gerardo2}:

\begin{align}
\sigma^{(f)}=
\left(
\begin{array}{cccc}
 \beta  & \zeta  & \ldots  & \zeta  \\
 \zeta  & \beta  & \zeta  & \vdots  \\
 \vdots  & \zeta  & \ddots & \zeta  \\
 \zeta  & \cdots  & \zeta  & \beta  \\
\end{array}
\right)
\mbox{.}
\end{align}
The elements of the covariance matrix depend on squeezing $r$ and number of modes $Z$ in the following way:

\begin{align}
\beta&=
\left(
\begin{array}{cc}
 b & 0 \\
 0 & b \\
\end{array}
\right)
 \mbox{~} \mbox{,~}
 \zeta=
 \left(
\begin{array}{cc}
 z_1 & 0 \\
 0 & z_2 \\
\end{array}
\right)
\mbox{,}
\\
z_1&=\frac{2 (Z-2) \sinh ^2(2 r)+Z \sinh (4 r)}{Z \sqrt{2 (Z-1) \cosh (4 r)+(Z-2) Z+2}}
\mbox{,}
\\
z_2&=\frac{2 (Z-2) \sinh ^2(2 r)-Z \sinh (4 r)}{Z \sqrt{2 (Z-1) \cosh (4 r)+(Z-2) Z+2}}
\mbox{,}
\\
b&=\frac{\sqrt{2 (Z-1) \cosh (4 r)+(Z-2) Z+2}}{Z}
\mbox{.}
\end{align}

Let us now investigate how the purity of the state changes, when observed by $Z$ symmetrically accelerated observers. The non-inertial effects are represented by the action of the Gaussian channel we have introduced above. A purity $\mu$ of a generic $Z$–mode Gaussian state described by a covariance matrix $\sigma$ can be written as \cite{purity}:

\begin{align}
\mu =\frac{1}{2^{Z}\sqrt{\det\sigma }},
\end{align}
therefore we will be interested in a relative purity of the state defined as:

\begin{align}
\mu_{rel}&=\frac{\mu^{(d)}}{\mu^{(f)}}
=
\sqrt{\frac{\det \sigma^{(f)}}{\det \sigma^{(d)}}}
\nonumber
\\
&=
\sqrt{\frac{\det \sigma^{(f)}}{\det
\left(           
M\bm{\sigma}^{(f)}M^T+N
\right)}}
\mbox{.}
\end{align}

The choise of modes, introduced in the previous section, leads to a significant simplification of the $M$ matrix. Due to the fact that we neglected all the $\beta$ coefficients and the values of $\alpha$ are real we have:

\begin{align}
M &=\left( \alpha \mathbb{I}_{2\times2} \right)^{\bigoplus Z}
\nonumber
\\
N &=\left[\left( 1-\alpha^2\right) \mathbb{I}_{2\times2} \right]^{\bigoplus Z}
\mbox{.}
\end{align}

\begin{figure}
\centering
\includegraphics[width=1\linewidth]{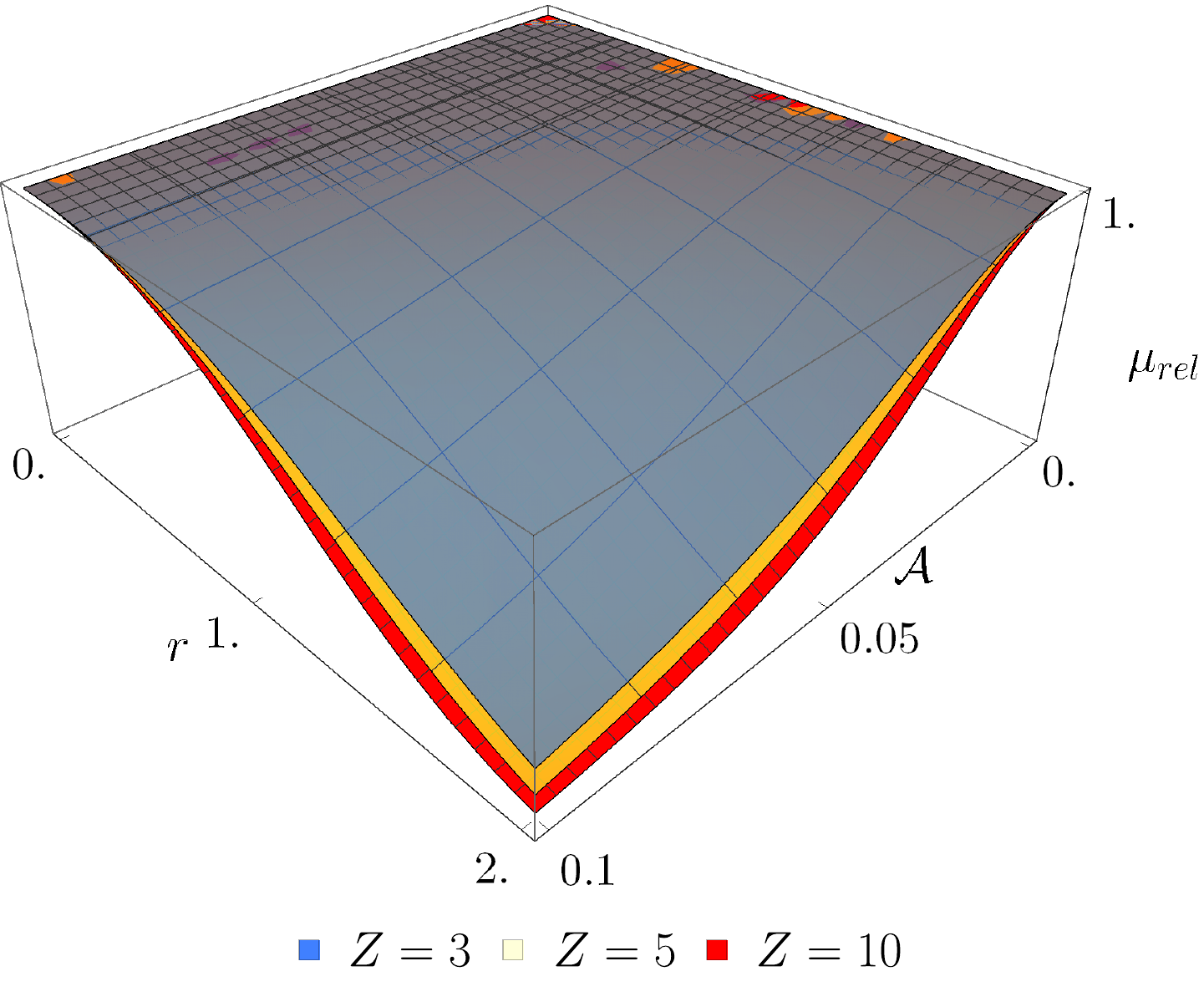}
\caption{The dependence of the $\mu_{rel}=\sqrt{\frac{\det \sigma^{(f)}}{\det \sigma^{(d)}}}$ on the observer's acceleration $\mathcal{A}$ and squeezing parameter $r$. The plot for different $Z$-number of modes. } 
\label{fig:purity}
\end{figure}

The resulting relative purity as a function of the proper acceleration ${\cal A}$ of all the $Z$ observers, as well as the initial level of squeezing $r$ of the symmetric Gaussian state has been studied numerically. The results are plotted for different values of $Z$ in Fig.~\ref{fig:purity}. We find that the purity of the $Z$-partite pure state is reduced with acceleration and the effect is stronger for the states with the larger entanglement. We also find that the degradation of purity is increased with the number $Z$ of parties involved.

\section{Conclusions\label{SecConcl}}
In this paper, we studied transformation of multiomode Gaussian states between an inertial frame and a uniformly accelerated frame of reference. We generalized the previous results \cite{Ahmadi} to a generic scheme for transforming an arbitrary, multimode Gaussian where each of the modes can be observed by a different accelerated observer. Finally, we applied our results to the exemplary case, where each of the observers accelerates in a different direction in space observing a fully symmetric $Z$-mode pure squeezed vacuum state. We showed how the purity of the initial state has changed when we had considered $Z$ symmetrically accelerated observers.

\begin{acknowledgments}
This work was supported by the National Science Centre,
Sonata BIS Grant No.~2012/07/E/ST2/01402.
\end{acknowledgments}

\onecolumngrid
\appendix
\section{Computing the covariance matrix of the Minkowski vacuum in the accelerated reference frame when $D=0$\label{vaccov}}
In this appendix we derive elements of the covariance matrix of the Minkowski vacuum state in the accelerated reference frame when $D=0$. Let us first compute the even, diagonal elements of the matrix $(\sigma_{\text{vac}}^{(d)} )_{2n,2n}$. Since the vacuum state has a vanishing mean value of all the annihilation and creation operators $\hat{d}_n^{(\dagger)}$, we have:
\begin{align}
\left(\sigma_{\text{vac}}^{(d)}\right)_{2n,2n}=
_{\text{M}}\bra{0}
\left\{  \frac{\hat{d}_n-\hat{d}_n^{\dagger}}{\sqrt{2}i}, 
\frac{\hat{d}_n-\hat{d}_n^{\dagger}}{\sqrt{2}i} \right\}\ket{0}_\text{M},
\nonumber
\end{align}
In order to proceed with the computation we will use some of the results derived in \cite{Ahmadi}, namely:
\begin{align}
_{\text{M}}\bra{0}\hat{b}_{\Lambda\Omega}\hat{b}_{\Lambda\Omega'}\ket{0}_{\text{M}}&=
_{\text{M}}\bra{0}\hat{b}_{\Lambda\Omega}^{\dagger}\hat{b}_{\Lambda\Omega'}^{\dagger}\ket{0}_{\text{M}}
\propto \int \mathrm{d}k \alpha_{\Omega k}^{(\Lambda)}\alpha_{\Omega' k}^{(\Lambda)} =0
\mbox{,}
\\
_{\text{M}}\bra{0}\hat{b}_{\Lambda\Omega}\hat{b}_{\Lambda\Omega'}^{\dagger}\ket{0}_{\text{M}}&=
\int \mathrm{d}k \alpha_{\Omega k}^{(\Lambda)*}\alpha_{\Omega' k}^{(\Lambda)} =
\frac{\delta\left(\Omega-\Omega'\right)}{1-e^{-\frac{2\pi\Omega}{a}}}
\mbox{,}
\\
_{\text{M}}\bra{0}\hat{b}_{\Lambda\Omega}^{\dagger}\hat{b}_{\Lambda\Omega'}\ket{0}_{\text{M}}&=
e^{-\frac{\pi (\Omega+\Omega')}{a}}\int \mathrm{d}k \alpha_{\Omega k}^{(\Lambda)}\alpha_{\Omega' k}^{(\Lambda)*} =
\frac{\delta\left(\Omega-\Omega'\right)}{e^{\frac{2\pi\Omega}{a}}-1}
\mbox{,}
\\
_{\text{M}}\bra{0}\hat{b}_{\text{I}\Omega}\hat{b}_{\text{II}\Omega'}\ket{0}_{\text{M}}&=
_{\text{M}}\bra{0}\hat{b}_{\text{I}\Omega}^{\dagger}\hat{b}_{\text{II}\Omega'}^{\dagger}\ket{0}_{\text{M}}=
e^{-\frac{\pi \Omega}{a}}\int \alpha_{\Omega k}^{(\text{I})}\alpha_{\Omega' k}^{(\text{I})*} \rm{d}k=
\frac{\delta\left(\Omega-\Omega'\right)}{2\sinh\left(\frac{\pi\Omega}{a}\right)}
\mbox{,}
\\
_{\text{M}}\bra{0}\hat{b}_{\text{I}\Omega}^{\dagger}\hat{b}_{\text{II}\Omega'}\ket{0}_{\text{M}}&=_{\text{M}}\bra{0}\hat{b}_{\text{I}\Omega}\hat{b}_{\text{II}\Omega'}^{\dagger}\ket{0}_{\text{M}}
\propto\int \mathrm{d}k \alpha_{\Omega k}^{(\text{I})}\alpha_{\Omega' k}^{(\text{I})} = 0
\mbox{,}
\end{align}
Upon the substitution, we have:

\begin{align}
_{\text{M}}\bra{0}\hat{d}_{n}\hat{d}_{k}\ket{0}_{\text{M}}=&
\int_{0}^{\infty}
\mathrm{d}\Omega
\frac{
\left(\psi_{n},w_{\text{I}\Omega}\right)
\left(\psi_{k},w_{\text{II}\Omega}\right)
+
\left(\psi_{n},w_{\text{II}\Omega}\right)
\left(\psi_{k},w_{\text{I}\Omega}\right)
}{2\sinh\left(\frac{\pi\Omega}{a}\right)}
\mbox{,}
\\
_{\text{M}}\bra{0}\hat{d}_{n}\hat{d}_{k}^{\dagger}\ket{0}_{\text{M}}=&
\int_{0}^{\infty}
\mathrm{d}\Omega
\frac{
\left(\psi_{n},w_{\text{I}\Omega}\right)
\left(\psi_{k},w_{\text{I}\Omega}\right)^*
+
\left(\psi_{n},w_{\text{II}\Omega}\right)
\left(\psi_{k},w_{\text{II}\Omega}\right)^*
}{{1-e^{-\frac{2\pi\Omega}{a}}}}
\mbox{.}
\end{align}

This leads to the following result:
\begin{align}
\left(\sigma_{\text{vac}}^{(d)}\right)_{2n,2n}&=
~_{\text{M}}\bra{0}\{\hat{d}_n,\hat{d}_n^{\dagger}\}\ket{0}_\text{M}-
_{\text{M}}\bra{0}\hat{d}_n\hat{d}_n+\hat{d}_n^{\dagger}\hat{d}_n^{\dagger}\ket{0}_\text{M}
\nonumber
\\
&=~_{\text{M}}\bra{0}\{\hat{d}_n,\hat{d}_n^{\dagger}\}\ket{0}_\text{M}-\int_{0}^{\infty}\mathrm{d}\Omega\frac{\operatorname{Re}\left[
\left(\psi_{n},\omega_{\text{I}\Omega}\right)\left(\psi_{n},\omega_{\text{II}\Omega}\right)\right] 
}{\sinh\left(\frac{\pi\Omega}{a}\right)}
\mbox{,}
\end{align}
but:
\begin{align}
_{\text{M}}\bra{0}\hat{d}_n\hat{d}_n^{\dagger}+\hat{d}_n^{\dagger}\hat{d}_n\ket{0}_\text{M}
=1+
\int_{0}^{\infty}
\mathrm{d}\Omega
\left(
|\left(\psi_{n},w_{\text{I}\Omega}\right)|^2+
|\left(\psi_{n},w_{\text{II}\Omega}\right)|^2
\right)
\frac{e^{-\frac{\pi\Omega}{a}}}{\sinh\left(\frac{\pi\Omega}{a}\right)}
\mbox{.}
\end{align}
Finally, an even diagonal element of the covariance matrix has the form:
\begin{align}
\left(\sigma_{\text{vac}}^{(d)}\right)_{2n,2n}=&
1+
\int_{0}^{\infty}
\mathrm{d}\Omega
\left(
|\left(\psi_{n},w_{\text{I}\Omega}\right)|^2+
|\left(\psi_{n},w_{\text{II}\Omega}\right)|^2
\right)
\frac{e^{-\frac{\pi\Omega}{a}}}{\sinh\left(\frac{\pi\Omega}{a}\right)}
\nonumber
\\
&-
\int_{0}^{\infty} \mathrm{d}\Omega \frac{\operatorname{Re}\left[
\left(\psi_{n},w_{\text{I}\Omega}\right)\left(\psi_{n},w_{\text{II}\Omega}\right)\right]
}{\sinh\left(\frac{\pi\Omega}{a}\right)}
\mbox{.}
\end{align}
For odd diagonal elements the derivation is analogous. We have:

\begin{align}
\left(\sigma_{\text{vac}}^{(d)}\right)_{2n-1,2n-1} =~&_{\text{M}}\bra{0}
2\left(\frac{\hat{d}_n+\hat{d}_n^{\dagger}}{\sqrt{2}}\right)^2\ket{0}_\text{M}
\nonumber
\\
=&
~_{\text{M}}\bra{0}\hat{d}_n\hat{d}_n+\hat{d}_n^{\dagger}\hat{d}_n^{\dagger}\ket{0}_\text{M}+
_{\text{M}}\bra{0}\hat{d}_n\hat{d}_n^{\dagger}+\hat{d}_n^{\dagger}\hat{d}_n\ket{0}_\text{M}
\nonumber
\\
=&\int_{0}^{\infty}
\mathrm{d}\Omega
\frac{\operatorname{Re}\left[
\left(\psi_{n},w_{\text{I}\Omega}\right)\left(\psi_{n},w_{\text{II}\Omega}\right)\right] 
}{\sinh\left(\frac{\pi\Omega}{a}\right)}
+
_{\text{M}}\bra{0}\hat{d}_n\hat{d}_n^{\dagger}+\hat{d}_n^{\dagger}\hat{d}_n\ket{0}_\text{M}\nonumber \\
=&1+
\int_{0}^{\infty}
\mathrm{d}\Omega
\left(
|\left(\psi_{n},w_{\text{I}\Omega}\right)|^2+
|\left(\psi_{n},w_{\text{II}\Omega}\right)|^2
\right)
\frac{e^{-\frac{\pi\Omega}{a}}}{\sinh\left(\frac{\pi\Omega}{a}\right)}
\\&+
\int_{0}^{\infty}
\mathrm{d}\Omega
\frac{\operatorname{Re}\left[
\left(\psi_{n},w_{\text{I}\Omega}\right)\left(\psi_{n},w_{\text{II}\Omega}\right)\right]
}{\sinh\left(\frac{\pi\Omega}{a}\right)}
\mbox{.}
\end{align}
Similarily, the nearest off-diagonal elements of the matrix have the following form:

\begin{align}
\left(\sigma_{\text{vac}}^{(d)}\right)_{2n,2n-1}
&=~_{\text{M}}\bra{0}
\{  \frac{\hat{d}_n+\hat{d}_n^{\dagger}}{\sqrt{2}}
, 
\frac{\hat{d}_n-\hat{d}_n^{\dagger}}{\sqrt{2}i}
\}\ket{0}_\text{M}=
-i ~_{\text{M}}\bra{0}
\hat{d}_n\hat{d}_n
-
\hat{d}_n^{\dagger}
\hat{d}_n^{\dagger}
\ket{0}_{\text{M}}=
\nonumber
\\
&=
\int_{0}^{\infty}
\mathrm{d}\Omega
\frac{\operatorname{Im}\left[
\left(\psi_{n},w_{\text{I}\Omega}\right)\left(\psi_{n},w_{\text{II}\Omega}\right)\right]
}{\sinh\left(\frac{\pi\Omega}{a}\right)}
\mbox{.}
\end{align}
All the remaining matrix elements are:

\begin{align}
\left(\sigma_{\text{vac}}^{(d)}\right)_{2n-1,2k-1}=&
~_{\text{M}}\bra{0}
\{  \frac{\hat{d}_n+\hat{d}_n^{\dagger}}{\sqrt{2}}, 
\frac{\hat{d}_k+\hat{d}_k^{\dagger}}{\sqrt{2}}\}\ket{0}_\text{M}\mbox{,}
\\
\left(\sigma_{\text{vac}}^{(d)}\right)_{2n-1,2k}=&
~_{\text{M}}\bra{0}
\{  \frac{\hat{d}_n+\hat{d}_n^{\dagger}}{\sqrt{2}}, 
\frac{\hat{d}_k-\hat{d}_k^{\dagger}}{\sqrt{2}i}\}\ket{0}_\text{M}
\mbox{,}
\\
\left(\sigma_{\text{vac}}^{(d)}\right)_{2n,2k-1}=&
~_{\text{M}}\bra{0}
\{  \frac{\hat{d}_n-\hat{d}_n^{\dagger}}{\sqrt{2}i}, 
\frac{\hat{d}_k+\hat{d}_k^{\dagger}}{\sqrt{2}}\}\ket{0}_\text{M}\mbox{,}
\\
\left(\sigma_{\text{vac}}^{(d)}\right)_{2n,2k}=&
~_{\text{M}}\bra{0}
\{  \frac{\hat{d}_n-\hat{d}_n^{\dagger}}{\sqrt{2}i}, 
\frac{\hat{d}_k-\hat{d}_k^{\dagger}}{\sqrt{2}i}\}\ket{0}_\text{M},
\end{align}
where $k\in \big\{n+1,n+2,\cdots,Z\big\}$. Following the same procedure as for the diagonal elements, we find:
\begin{align}
\left(\sigma_{\text{vac}}^{(d)}\right)_{2n-1,2k-1}
=&
_{\text{M}}\bra{0}
\hat{d}_{n}\hat{d}_{k}
+
\left(\hat{d}_{n}\hat{d}_{k}^{\dagger}\right)^*
+
\hat{d}_{n}\hat{d}_{k}^{\dagger}
+
\left(\hat{d}_{n}\hat{d}_{k}\right)^*
\ket{0}_{\text{M}}
\nonumber
\\
=&
2\operatorname{Re}
\left(
_{\text{M}}\bra{0}
\hat{d}_{n}\hat{d}_{k}
\ket{0}_{\text{M}}
\right)
+
2\operatorname{Re}
\left(
_{\text{M}}\bra{0}
\hat{d}_{n}\hat{d}_{k}^{\dagger}
\ket{0}_{\text{M}}
\right)
\nonumber
\\
=&\int_{0}^{\infty}
\mathrm{d}\Omega
\Big(
\frac{
\operatorname{Re}
\left[
\left(\psi_{n},w_{\text{I}\Omega}\right)
\left(\psi_{k},w_{\text{II}\Omega}\right)
+
\left(\psi_{n},w_{\text{II}\Omega}\right)
\left(\psi_{k},w_{\text{I}\Omega}\right)
\right]
}{\sinh\left(\frac{\pi\Omega}{a}\right)}
\nonumber
\\
&+
\frac{
e^{\frac{\pi\Omega}{a}}
\operatorname{Re}
\left[
\left(\psi_{n},w_{\text{I}\Omega}\right)
\left(\psi_{k},w_{\text{I}\Omega}\right)^*
+
\left(\psi_{n},w_{\text{II}\Omega}\right)
\left(\psi_{k},w_{\text{II}\Omega}\right)^*
\right]
}{\sinh\left(\frac{\pi\Omega}{a}\right)}
\Big)
\mbox{.}
\end{align}
and analogously for the remaining elements.

\section{Computing the covariance matrix of the Minkowski vacuum in the accelerated reference frame when $D\neq0$\label{vaccovd}}
In this appendix we generalize the results obtained in Appendix \ref{vaccov} for the case, when $D\neq 0$.
From the equation \eqref{eqn:10} we have for $\Lambda\in\big\{\text{I},\text{II}\big\}$:
\begin{align}
\hat{b}_{\Lambda\Omega}=
\int 
\mathrm{d}k
\left(
\left( w_{\Lambda\Omega},u_k\right) \hat{a}_k
+
\left( w_{\Lambda\Omega},u_k^*\right) \hat{a}_k^{\dagger}
\right)
+
\left(w_{\Lambda\Omega},\Phi_{\text{III}}(D)\right)
\mbox{,}
\end{align}
but
$\left(w_{\Lambda\Omega},\Phi_{\text{III}}(D)\right)=0$ because decomposition of $\Phi_{\text{III}}(D)$ contains only modes from region between Rindler wegdes. Therefore:
\begin{align}
\hat{b}_{\Lambda\Omega}=
\int 
\mathrm{d}k
\left(
\alpha_{\Omega k}^{(\Lambda)*}\hat{a}_k
-\beta_{\Omega k}^{(\Lambda)*}\hat{a}_k^{\dagger}
\right)
\mbox{,}
\end{align}
where the Bogolyubov transformation coefficients are defined as:
\begin{align}
\alpha_{\Omega k}^{(\Lambda)}&=\left(u_k,w_{\Lambda\Omega}\right),
\\
\beta_{\Omega k}^{(\Lambda)}&=-\left(u_k^*,w_{\Xi\Omega}\right).
\end{align}
 A transition from the case of $D=0$ to the case when $D\neq 0$ corresponds to the following change of the Bogolyubov coefficients \cite{Ahmadi}:

\begin{align}
\tilde{\alpha}_{\Omega k}^{(\text{I})}&\rightarrow e^{-i\frac{D}{2}k}\alpha_{\Omega k}^{(\text{I})}
\mbox{,}
\\
\tilde{\alpha}_{\Omega k}^{(\text{II})}&\rightarrow e^{i\frac{D}{2}k}\alpha_{\Omega k}^{(\text{II})}
\mbox{,}
\\
\tilde{\beta}_{\Omega k}^{(\text{I})}&\rightarrow e^{i\frac{D}{2}k}\beta_{\Omega k}^{(\text{I})}
\mbox{,}
\\
\tilde{\beta}_{\Omega k}^{(\text{II})}&\rightarrow e^{-i\frac{D}{2}k}\beta_{\Omega k}^{(\text{II})}
\mbox{.}
\end{align}

Therefore in the case of $D\neq 0$ we have:
\begin{align}
_{\text{M}}\bra{0}\hat{b}_{\text{I}\Omega}^{\dagger}\hat{b}_{\text{II}\Omega'}\ket{0}_{\text{M}}
=&
_{\text{M}}\bra{0}
\int
\mathrm{d}k
\left(
 \tilde{\alpha}_{\Omega k}^{(\text{I})}\hat{a}_k^{\dagger}
-\tilde{\beta}_{\Omega k}^{(\text{I})}\hat{a}_k
\right)
\int 
\mathrm{d}k'
\left(
\tilde{\alpha}_{\Omega' k'}^{(\text{II})*}\hat{a}_{k'}
-\tilde{\beta}_{\Omega' k'}^{(\text{II})*}\hat{a}_{k'}^{\dagger}
\right)
\ket{0}_{\text{M}}=
\nonumber
\\
=&
\int 
\mathrm{d}k
\tilde{\beta}_{\Omega k}^{(\text{I})}\tilde{\beta}_{\Omega' k}^{(\text{II})*}
=
\int
\mathrm{d}k
 e^{i\frac{D}{2}k} \beta_{\Omega k}^{(\text{I})}e^{i\frac{D}{2}k}\beta_{\Omega' k}^{(\text{II})*} 
=
\int 
\mathrm{d}k
e^{iDk} e^{-\frac{\pi\Omega}{a}}\alpha_{\Omega k}^{(\text{I})}
e^{-\frac{\pi\Omega'}{a}}\alpha_{\Omega' k}^{(\text{I})} =
\nonumber
\\
=&e^{-\frac{\pi(\Omega+\Omega')}{a}}
\int 
\mathrm{d}k
e^{iDk}\alpha_{\Omega k}^{(I)}\alpha_{\Omega' k}^{(\text{I})} 
\mbox{.}
\end{align}
An analogous procedure involving other quadratic mononomials of the annihilation and creation operators $ \hat{b}_n^{(\dagger)} $ and $ \hat{b}_k^{(\dagger)}$ leads to the following results:

\begin{align}
_{\text{M}}\bra{0}\hat{b}_{\Xi\Omega}\hat{b}_{\Xi\Omega'}\ket{0}_{\text{M}}&=
_{\text{M}}\bra{0}\hat{b}_{\Xi\Omega}^{\dagger}\hat{b}_{\Xi\Omega'}^{\dagger}\ket{0}_{\text{M}}=0
\mbox{,}
\\
_{\text{M}}\bra{0}\hat{b}_{\Xi\Omega}\hat{b}_{\Xi\Omega'}^{\dagger}\ket{0}_{\text{M}}&=\frac{\delta\left(\Omega-\Omega'\right)}{1-e^{-\frac{2\pi\Omega}{a}}}
\mbox{,}
\\
_{\text{M}}\bra{0}\hat{b}_{\Xi\Omega}^{\dagger}\hat{b}_{\Xi\Omega'}\ket{0}_{\text{M}}&=\frac{\delta\left(\Omega-\Omega'\right)}{e^{\frac{2\pi\Omega}{a}}-1}
\mbox{,}
\\
_{\text{M}}\bra{0}\hat{b}_{\text{I}\Omega}\hat{b}_{\text{II}\Omega'}\ket{0}_{\text{M}}&=I_{1}(\Omega,\Omega')
\mbox{,}
\\
_{\text{M}}\bra{0}\hat{b}_{\text{II}\Omega'}\hat{b}_{\text{I}\Omega}\ket{0}_{\text{M}}
&=I_{1}(\Omega',\Omega)
\mbox{,}
\\
_{\text{M}}\bra{0}\hat{b}_{\text{I}\Omega}^{\dagger}\hat{b}_{\text{II}\Omega'}^{\dagger}\ket{0}_{\text{M}}&= I_{1}(\Omega',\Omega)
\mbox{,}
\\
_{\text{M}}\bra{0}\hat{b}_{\text{II}\Omega'}^{\dagger}\hat{b}_{\text{I}\Omega}^{\dagger}\ket{0}_{\text{M}}&=
I_{1}(\Omega,\Omega')
\mbox{,}
\\
_{\text{M}}\bra{0}\hat{b}_{\text{I}\Omega}^{\dagger}\hat{b}_{\text{II}\Omega'}\ket{0}_{\text{M}}&=I_3(\Omega,\Omega')e^{\frac{\pi(\Omega+\Omega')}{a}(1-\frac{D}{|D|})}
\mbox{,}
\\
_{\text{M}}\bra{0}\hat{b}_{\text{II}\Omega'}\hat{b}_{\text{I}\Omega}^{\dagger}\ket{0}_{\text{M}}
&=I_{3}(\Omega,\Omega')
\mbox{,}
\\
_{\text{M}}\bra{0}\hat{b}_{\text{I}\Omega}\hat{b}_{\text{II}\Omega'}^{\dagger}\ket{0}_{\text{M}}&=
I_{3}(\Omega,\Omega')
\mbox{,}
\\
_{\text{M}}\bra{0}\hat{b}_{\text{II}\Omega'}^{\dagger}\hat{b}_{\text{I}\Omega}\ket{0}_{\text{M}}&=
I_3(\Omega,\Omega')e^{\frac{\pi(\Omega+\Omega')}{a}(1-\frac{D}{|D|})},
\end{align}
where:

\begin{align}
I_1(\Omega,\Omega')&\equiv
\frac{
e^{\frac{\pi(\Omega-\Omega')}{2a}}
}{
2\pi a
\sqrt{
\sinh\frac{\pi\Omega}{a}
\sinh\frac{\pi\Omega'}{a}
}
}
e^{-\frac{D}{|D|}\frac{\pi(\Omega-\Omega')}{2a}}
K_{i\frac{\Omega-\Omega'}{a}}(|mD|)
\mbox{,}
\\
I_3(\Omega',\Omega)&\equiv
\frac{
e^{\frac{\pi(\Omega+\Omega')}{2a}}
}{
2\pi a
\sqrt{
\sinh\frac{\pi\Omega}{a}
\sinh\frac{\pi\Omega'}{a}
}
}
e^{-\frac{D}{|D|}\frac{\pi(\Omega+\Omega')}{2a}}
K_{i\frac{\Omega+\Omega'}{a}}(|mD|)
\equiv
I_3(\Omega,\Omega')
\mbox{.}
\\
I_2(\Omega,\Omega')
&\equiv
\frac{
e^{-\frac{\pi(\Omega+\Omega')}{2a}}
}{
4\pi a \sqrt{
\sinh\frac{\pi\Omega}{a}
\sinh\frac{\pi\Omega'}{a}
}
}
\int
\mathrm{d}k
\frac{e^{iDk}}{\omega_k}
\left(
\frac{\omega_k+k}{\omega_k-k}
\right)^{-\frac{i(\Omega+\Omega')}{2a}}
\nonumber
\\&=
\frac{
e^{-\frac{\pi(\Omega+\Omega')}{2a}}
}{
4\pi a
\sqrt{
\sinh\frac{\pi\Omega}{a}
\sinh\frac{\pi\Omega'}{a}
}
}
e^{\frac{D}{|D|}\frac{\pi(\Omega+\Omega')}{2a}}
K_{i\frac{\Omega+\Omega'}{a}}(|mD|)
\nonumber
\\
&=
I_3(\Omega,\Omega')
e^{-\frac{\pi(\Omega+\Omega')}{a}}
e^{\frac{D}{|D|}\frac{\pi(\Omega+\Omega')}{a}}
\equiv
I_3(\Omega,\Omega')e^{\frac{\pi(\Omega+\Omega')}{a}(1-\frac{D}{|D|})}
\mbox{.}
\end{align}

It is worth noticing that the integral expressions $ I_1 $, $ I_2 $ and $ I_3 $ are real-valued and we can omit complex conjugations. Moreover, the operators $ \hat{d}_n$ with different indices always commute. This leads to the following results:

\begin{align}
_{\text{M}}\bra{0}\hat{d}_{n}\hat{d}_{k}\ket{0}_{\text{M}}=&
\int_{0}^{\infty}\int_{0}^{\infty}
\mathrm{d}\Omega \mathrm{d}\Omega'
I_1(\Omega,\Omega')
[
\left(\psi_{n},w_{\text{I}\Omega}\right)\left(\psi_{k},w_{\text{II}\Omega'}\right)
+
\left(\psi_{n},w_{\text{II}\Omega}\right)\left(\psi_{k},w_{\text{I}\Omega'}\right)
]
\mbox{,}
\\
_{\text{M}}\bra{0}\hat{d}_{n}\hat{d}_{k}^{\dagger}\ket{0}_{\text{M}}=&
\int_{0}^{\infty}
\mathrm{d}\Omega
\frac{
\left(\psi_{n},w_{\text{I}\Omega}\right)
\left(\psi_{k},w_{\text{I}\Omega}\right)^*
+
\left(\psi_{n},w_{\text{II}\Omega}\right)
\left(\psi_{k},w_{\text{II}\Omega}\right)^*
}{1-e^{-\frac{2\pi\Omega}{a}}} 
\nonumber
\\
&+
\int_{0}^{\infty} \int_{0}^{\infty}
\mathrm{d}\Omega \mathrm{d}\Omega'
I_3(\Omega,\Omega')
\left[
\left(\psi_{n},w_{\text{I}\Omega}\right)
\left(\psi_{k},w_{\text{II}\Omega'}\right)^*
+
\left(\psi_{n},w_{\text{II}\Omega}\right)
\left(\psi_{k},w_{\text{I}\Omega'}\right)^*
\right].
\end{align}

The above expressions are sufficient to completely characterize all the elements of the covariance matrix in the case of $D\neq 0$.

\end{document}